\documentclass[12pt]{article}
\usepackage{latexsym}
\usepackage{epsfig}
\usepackage{graphicx}
\usepackage[english]{babel}
\usepackage{epstopdf}
\newcommand \jp {$J/\psi \,$}

\newcommand{\beq}{\begin{equation}}
\newcommand{\eeq}{\end{equation}}
\newcommand{\be}{\begin{eqnarray}}
\newcommand{\ee}{\end{eqnarray}}

\begin{document}
\title{The LHC potential for study of the small $x$ gluon physics
 in ultraperipheral collisions of 3.5 TeV protons.}
\author {
V.~Rebyakova\\
\it Petersburg State Technical University\\
\it  St.Petersburg, Russia\\
M. ~Strikman\\
\it Department of Physics,
Pennsylvania State University,\\
\it University Park, Pennsylvania  16802, USA
\\
M.~Zhalov\\
\it Petersburg Nuclear Physics Institute, Gatchina, 
Russia}

\date{}
\maketitle
\begin{abstract}
 We argue that already the first year LHC run at $\sqrt{s}={\mbox 7 TeV}$
 will provide a quick and  effective way to test
 the dynamics of the 
color dipole - gluon interactions and the small $x$
behavior of the gluon density in the proton by 
 studying   vector meson photoproduction in ultraperipheral proton-proton collisions.
\end{abstract}

It is expected now that during 2010 the LHC will operate 
 at energies of the colliding proton beams of  3.5 TeV and 
luminosity $L_{pp}\approx 10^{32} cm^{-2} s^{-1}$. 
This raises the question whether it will be possible 
to realize part of the program of small $x$ gluon physics in the ultraperipheriral collisions 
\cite{Baltz:2007kq} already during the first year of the operation of the LHC. 
In this paper we suggest to study large momentum transfer
vector meson photoproduction with target dissociation and coherent $J/\psi$ photoproduction
in ultraperipheral proton-proton collisions (UPC) at $\sqrt s\approx 7$ TeV.
We show that  this would allow one
to investigate the behavior of the gluon density in a proton at small $x$
down to $x\approx 10^{-5}$,
as well as  
to determine the energy dependence of the color dipole - gluon
elastic scattering amplitude.

Understanding of the small $x \le 10^{-3}$ dynamics of gluon interactions at moderate 
virtualities  is important for a realistic description of the $pp$ 
collisions at the LHC energies. 
The small x processes are a  topic of numerous theoretical studies. 
Currently most of the  experimental information about such processes comes from 
experiments which were performed at the electron-proton HERA collider. 
It would be natural to extend these studies at the LHC. 
A possible solution  is to use the so-called ultraperipheral collisions (UPC).
 So far the main focus of the theoretical UPC studies (for a recent review see \cite{Baltz:2007kq}) 
was on the heavy ion and proton - ion collisions. Notable exceptions \cite{pplhctheor},
\cite {Aaltonen:2009kg} are
the theoretical predictions  of the exclusive \jp production in $pp$ collisions
at $\sqrt{s}=14$ TeV and theoretical and experimental studies in
$p\bar p$ collisions at energies of Tevatron.

Recently we  pointed out that it would be possible to look for nonlinear effects in 
the interaction of small dipoles with strong gluon fields by studying the 
large momentum transfer 
$-t\equiv (p_{\gamma} - p_{J/\psi})^2$ process:
$\gamma + T \to J/\psi + {\rm  \, gap}  + X$
in the UPC of heavy ions with proton and nuclear target \cite{Frankfurt:2006tp},
\cite{Frankfurt:2008er},
\cite{Frankfurt:2008et}.
Here we want to explore the feasibility of using the same UPC probe for  
pp
scattering at the LHC. In practical  terms it has a number of appealing features: 
(i) the ability to vary $t$ allows one to change the resolution scale by a factor 
$\sim Q_0^2(1-t/4m_c^2)$, where $m_c$ is the charm quark mass and 
$Q_0^2  \sim 2.5 \div 3.5 \mbox{GeV}^2$ is the 
scale probed in the exclusive photoproduction of \jp; 
(ii) the possibility to use triggers involving hadron production.
 Such measurements will also serve as a benchmark 
for  similar measurements in the heavy ion collisions where this process provides an effective
 way to probe propagation of the small dipoles through the strong gluon fields.
 
 The cross section of the vector meson photoproduction in the proton-proton UPC 
is dominated by scattering at large impact parameters $b>2R_p$ corresponding
to the transverse momentum transferred through photon $p_t <0.1$ GeV. Hence
it is reasonable to
expect that
soft initial and final state inelastic $pp$ interactions are still not too significant.
Numerically 
a reduction factor due to these rescatterings 
should be of the order of 0.8 as given by the estimates in the 
literature for exclusive \jp production in the pp UPC \cite{pplhctheor}. We
neglect
 these absorption effects in the  current calculation.  
 Then, the cross section for  the UPC process $p+p\to p+\gamma+p\to p+V+X$ in the first 
approximation can be written as a convolution
of the photon flux from the fast moving proton and the cross section of the large
$t$ and rapidity gap vector meson (V) photoproduction off the proton target
\begin{equation}
{\frac {d\sigma_{pp\rightarrow p+X+V}} {dxdtdy}}=
{\frac {dN_{\gamma /p} (y)} {dy}} \cdot {\frac {d\sigma_{\gamma +p \rightarrow X+V}(x,y,t)} {dxdt}}.
\end{equation}
 The flux of the photons with momentum $k$ from the fast moving proton was 
calculated in \cite{Drees:1988pp}.

The process   $\gamma +p\to$V + "gap" + X  belongs to a class of reactions 
where the selection of large $-t\gg 1/r_N^2$ ensures that 
the dipole quark-antiquark component of
the photon wave function which transforms into vector meson has the small size 
$r_{V}\approx 1/\sqrt{-t}$.
 Besides, the important consequence of such a selection is that
the transverse momenta 
remain large in all rungs of the gluon ladder and therefore 
two gluons are attached  to one parton in the target.
 An attachments of the ladder to two different partons of the proton is  suppressed by 
a power of $t$. 
As a result the  cross section can be written in a factorized form as a product 
of the gluon density in the 
proton $g_{p}(x)$ (with a small correction for scattering off the 
quarks which we will not write
down
explicitly)  
and the elastic cross section 
of  scattering of a small dipole off a gluon\cite{ryskin}:
\begin{eqnarray}
\frac {d\sigma_{\gamma p\to V X}} {dt d{x}}=
\frac {d\sigma_{\gamma g \to V g}} {dt} g_{p}(x,t). 
\label{DGLAPBFKL}
\end{eqnarray}
Here
fraction $x$ of the target
proton momentum which carries the parton interacting with the dipole is
\begin{equation}
{x}=-t/(M_X^2 - m_N^2 -t),
\label{xval}
\end{equation}
 and $M_X$ 
is the invariant mass of the produced hadron system. 

 Experimentally it is rather difficult 
to measure $M_X$ directly but at fixed $t$ and $M^2_{X}>>-t>>m^2_N$ 
it can be expressed with a good accuracy 
through the rapidity interval , $\Delta y$,  occupied by the system $X$
\begin{eqnarray} 
\Delta y =\ln {\biggl (\frac {M^2_{X}} {m_N\sqrt{-t}}\biggr )}=
\ln {\biggl (\frac {\sqrt{-t}} {xm_N}\biggr )}.
\label{gapdef}
\end{eqnarray}
Hence, accurate measurement of $\Delta y$ ensures determination of $M_X$ and the value 
of $x$ of the interacting gluon.

Theoretical analysis  of the elastic dipole-gluon scattering through the exchange by the gluon ladder
leads
in the leading and next-to-leading log approximations to the 
amplitude which  increases with
energy at fixed $t$ as a power of $s$, $f(s,t)\propto (s_{\gamma g} /|t|)^{\alpha(t)-1}$, where
$s_{\gamma g}=xs_{\gamma N}$.
 The parameter $\alpha(t)=\alpha (0)+\alpha ^\prime t$ can be treated as
a  trajectory of the hard
effective pomeron responsible for the elastic dipole-gluon scattering. 
Calculations of intercept $\alpha(0)$ in the BFKL approach give different estimates ranging 
from value $\alpha(0)\approx 1.4$ in leading order BFKL to $\alpha(0)\approx 1.1$ in 
next-to-leading order and intermediate
value $\alpha(0)\approx 1.2$ in the resummed next-to-leading order
 BFKL; for a review see \cite{Colferai}. In our analysis
of the HERA data for the large $t$
and rapidity gap process $\gamma +p\to J/\psi+X$
we were trying to extract the value of 
$\alpha(0)$
from fitting  the data with the following parametrization\cite{Frankfurt:2008er}  
\begin{eqnarray}
{\frac {d\sigma_{\gamma +g\to J/\psi +g}} {dt}}=
\frac {C} {( t_{0}-t)
(M_{J/\psi}^2-t)^{3}}
\cdot \biggl [ {\frac {xW_{\gamma p}^2} {\sqrt{(t_{0} -t)
(M_{J/\psi}^2 -t})}}\biggr ]^{2(\alpha (t)-1)}. 
\label{gjpsi}
\end{eqnarray}
We fixed the scale $t_0=1\,GeV^2$ and
the quantities $C$, $\alpha(0)$ and $\alpha^\prime$ were used as free parameters.
A rather narrow energy interval $ 80\, GeV<W_{\gamma p}<200 \, GeV$ studied in the HERA experiments
and the experimental cuts  imposed due to 
 specific acceptances of the detectors 
do not allow one to perform 
stringent tests of the factorization approximation and 
result in a very modest 
sensitivity to the energy dependence of the dipole - gluon elastic cross section.
In particular, we found \cite{Frankfurt:2008er} that the data restrict the value of the slope parameters 
$\alpha^{\prime}\approx 0.005\div 0.01$ but allow change of the intercept
 $\alpha (0)$ in the range from 1.0 to 1.2.
Similar results have been obtained in \cite{Frankfurt:2006wg} where we
analyzed with the same model $\rho$ meson photoproduction (for other recent data analyses and  
references
see Refs.~\cite{Ivanov:2004ax}, \cite{Forshaw}). 

Note that in the HERA experiments the gap $\delta y$ between produced mass and $J/\psi$
was only 2-3 rapidity units that is obviously insufficient for developing
the BFKL dynamics (emitting of one additional gluon in the ladder requires at least 
two units of rapidity). Simple estimates show that at different LHC detectors  
it is possible to study this process with $\delta y\approx 2\div 6$. Hence,
exploring the ultraperipheral pp or ion-ion collisions at LHC significantly extends
the available kinematical region and can provide the capability to obtain more
detailed and precise information about the physics of the hard effective pomeron.

We suggest to explore in the LHC experiments two different options for 
studying the large momentum transfer and rapidity gap vector meson
 photoproduction in the proton-proton UPC.

 First, one can fix $\Delta y$ that
corresponds to fixed $M_X$ and, hence, fixed $x$ of the target gluon participating in the process and
study the cross section as a function of the rapidity of the vector meson. This kinematics allows one 
to investigate the energy dependence of the dipole-gluon elastic scattering through the exchange 
by the gluon ladder provided that the rapidity gap between the vector meson and produced mass $M_X$ is 
large enough. 
If a detector has sufficiently good acceptance it is possible 
 to increase statistics by summing
events with the produced mass $M_{X}\leq  M^{max}_{X}$. 
In theoretical estimates this procedure corresponds to the integration of the cross section
over $x$ in the range $x_{min}<x<1$ where $x_{min}$ can be found
from Eq. (\ref{xval}) using the value of $M_X^{max}$. Since the integral
$$I(x_{min})=\int \limits_{x_{min}}^{1} x^{2\alpha (t)-2}g_{p}(x)dx$$ grows 
with an increase of $M_X^{max}$
(decrease of $x_{min}$) due to the growth of the gluon density at small $x$, the choice of larger
$M_X^{max}$ will also increase the counting rate.
 However, because of the 
increasing uncertainties in the gluon density distributions with decrease of $x$, larger
$M_X^{max}$ leads to larger uncertainties in the analysis of the data. In our calculations
we used $M_X^{max}=25$ GeV that corresponds to $x_{min}\approx 0.005$ at $-t=3$ GeV$^2$.
\begin{figure}
\begin{center}
\epsfig{file=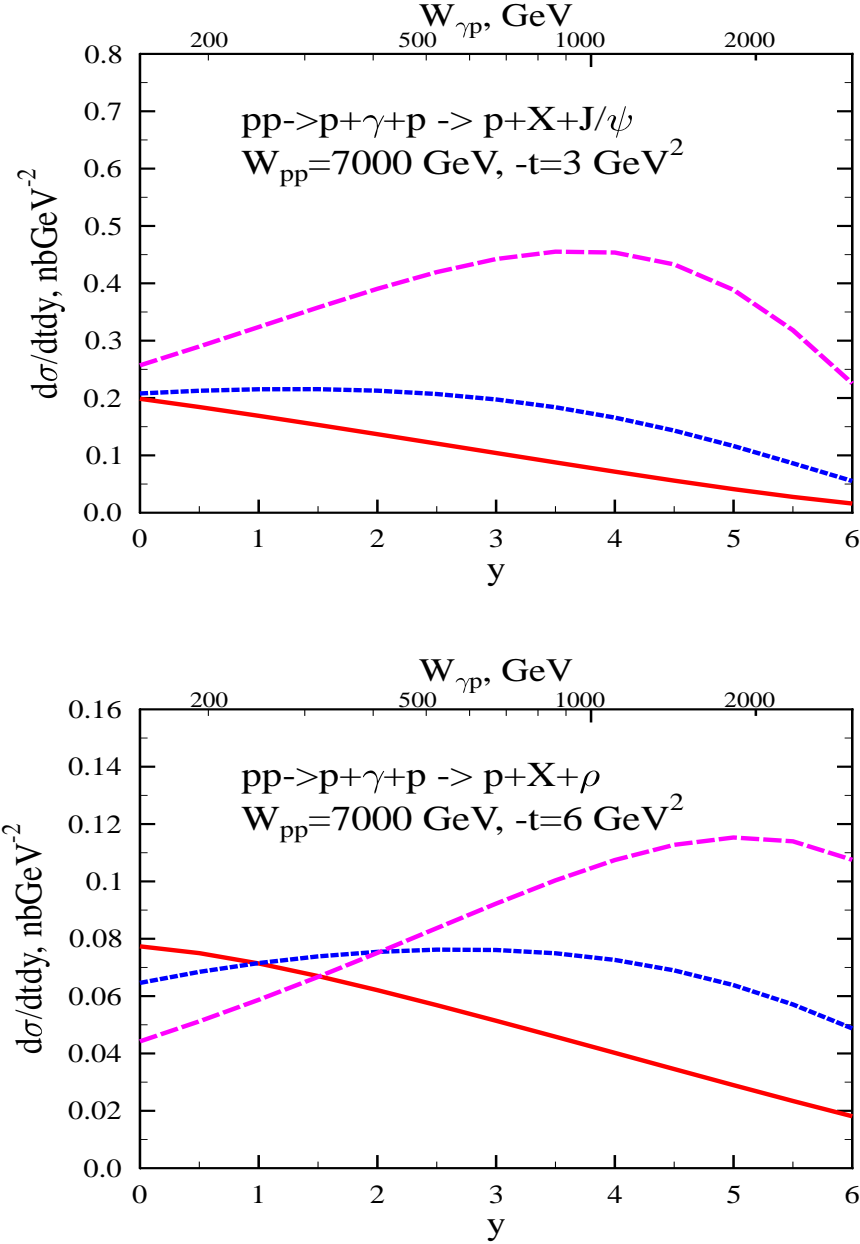, height=6.5in,width=4.5in}
 \caption{Rapidity distributions for the large $t$ $J/\psi$ and $\rho$ meson photoproduction in
 the ultraperipheral proton-proton collisions at LHC at $\sqrt{s}=7$ TeV.
Solid line: calculations with $\alpha(0)=1.0$; short dashed line: $\alpha(0)=1.1$; 
long dashed line: $\alpha(0)=1.2$. At rapidity of vector meson $y=0$ the rapidity gap between
vector meson and produced system $X$ is $\delta y=3$. }
 \label{endep}
\end{center}
\end{figure}
 In the interval $0.005\leq x\leq 1$ all modern sets of the gluon distributions give
close results, so, we calculated the cross sections in the approximation described 
above with gluon density
distribution given by CTEQ6L \cite{cteq6}. The results are presented  
in Fig.{\ref{endep} for the production of $J/\psi$ at $t=-3$ GeV$^2$ 
and of the $\rho$ meson at $-t=6$ GeV$^2$.
In the case of $\rho$ meson we choose larger $t$ in order to have large enough scale justifying
the perturbative QCD description.
 The standard method of $J/\psi$
detection is through the dilepton decays; hence, to estimate the counting rate for $J/\psi$ 
one has to account for
the branching factor $6\cdot 10^{-2}$. We estimate that 
it will be possible to collect tens of thousands of $J/\psi \to \mu^{+}\mu^{-}$ 
events in 1 year of running.
A more accurate estimate requires accounting for acceptance and other specifics of the detectors.
In spite of the decrease of cross section 
due to increase of $t$ ( asymptotically the cross section is scaled as $t^{-4}$) the number of events
in measurements of $\rho$ photoproduction can be significantly larger if the detector is capable 
of registering pions from $\rho$ decay in the forward direction.

Our calculations indicate that the ultraperipheral collisions 
of 3.5 TeV protons at the LHC will allow one to 
explore the region of $W_{\gamma p}$ from 100 GeV up to $ \approx 1000\div 2000$ GeV. Note,  
at HERA such measurements were restricted to 
 $80 \mbox{GeV} < W_{\gamma p}< 200 $ GeV.
Hence it appears that the discussed  measurements  will allow 
one to determine  the value of
$\alpha(0)$  with a good precision.

\begin{figure}
\begin{center}
\epsfig{file=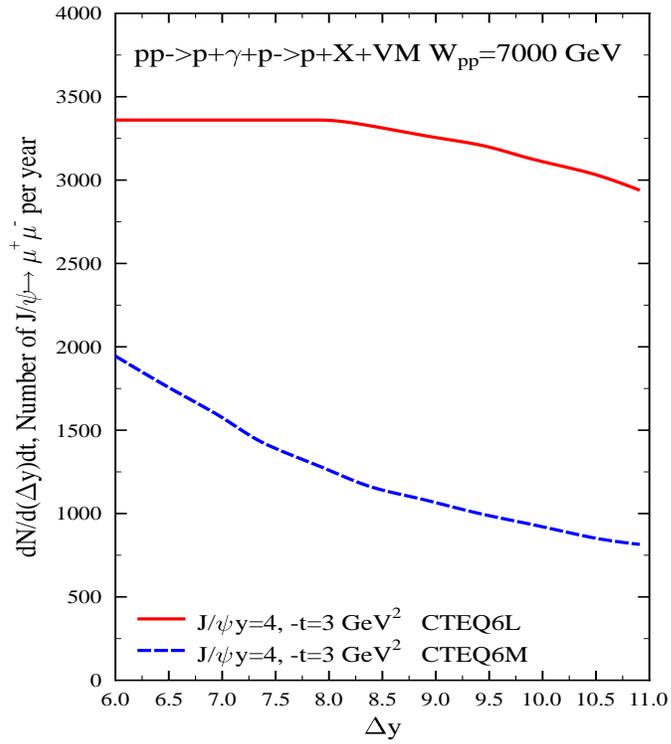, height=5.in,width=4.in}
 \caption{The large rapidity gap $J/\psi$
photoproduction in
the ultraperipheral proton-proton collisions   at $\sqrt{s}=7$ TeV
- number of $J/\psi \to \mu^{+}\mu^{-}$ events which could be accumulated
for the 1 year running period at luminosity $10^{32}\, cm^{-2}s^{-1}$.
}
 \label{xdep}
\end{center}
\end{figure}

The second 
possible strategy 
is to fix the rapidity of the vector meson 
and momentum transfer $t$
and increase 
the mass $M_X$ of the system $X$ produced  by target dissociation. 
In this case the energy of the dipole-gluon
collision remains 
 practically constant, while the rapidity interval occupied by the produced 
system $X$ grows with an increase of $M_X$ - a decrease of $x$ at fixed $t$, 
$x\approx \frac {-t} {M^2_X}$. 
Hence  such measurements allow  to study
the behavior of the nucleon gluon density at small $x$ at the  scale given by the value of $t$. 
 
We performed calculations with
the value of intercept $\alpha(0)=1.1$ and considered photoproduction of $J/\psi$ with
fixed rapidity $y=4$ at the momentum transfer $-t=3$ GeV$^2$. 
The number of $J/\psi \to \mu^{+} \mu^{-}$ events 
as
a function of $\Delta y$ for a 1 year running period at luminosity $10^{32}\, cm^{-2}s^{-1}$ is shown
in Fig.\ref{xdep} for gluon density distributions CTEQ6L and CTEQ6M.
The interval of change of $M_X$ from
25 GeV to 350 GeV corresponds to scanning of the gluon density in the proton in the range
$5\cdot 10^{-3}> x > 3\cdot 10^{-5}$. 
Obviously, accuracy of such 
scanning of the gluon density  essentially depends on the capability of the detector to
measure with good precision the value of $t$, say with $\Delta t \le 1$ GeV$^2$, and
the interval of rapidities $\Delta y$ occupied by the difractive produced system $X$
which determines
$x$ of the gluon. 
Also, to reach the region of very small $x<0.0001$ still keeping the reasonable value of
rapidity gap ($\delta y \ge 2$) one needs to detect the forward $J/\psi$ with the very high energy.
In principle this is possible with the forward muon spectrometer of the ALICE detector but
the acceptance will probably be small.

It seems that more preferable will be the study of 
the small x behavior of the gluon density in the proton 
from the 
measurement of the coherent $J/\psi$ photoproduction in the proton-proton UPC. 
Feasibility of such measurements has recently been demonstrated by CDF
Collaboration at energies of Tevatron \cite {Aaltonen:2009kg}.
The
cross sections for this process have been also predicted for the 
energy $\sqrt{s}=14$ TeV in \cite{pplhctheor}.
Here we demonstrate that the reasonable statistics can be accumulated 
already during the first year
running of the LHC at the energy $\sqrt{s}=7$ TeV.

\begin{figure}
\begin{center}
\epsfig{file=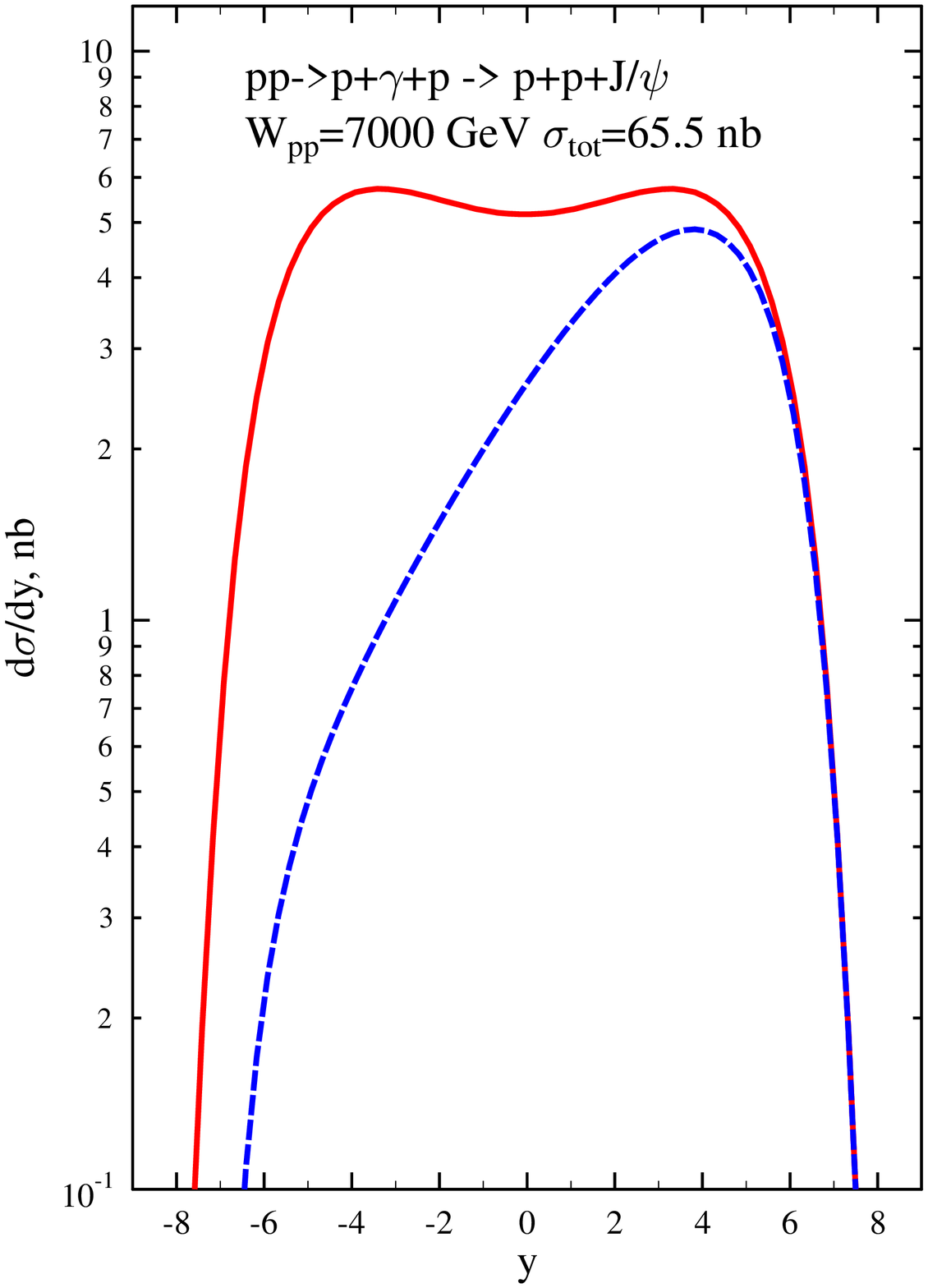, height=5.in,width=4.in}
 \caption{Rapidity distributions for the coherent $J/\psi$ photoproduction in
 the ultraperipheral proton-proton collisions at LHC at $\sqrt{s}=7$ TeV. 
 The dashed curve shows one-side contribution of $J/\psi$ photoproduction off the proton target.}
 \label{psicoh}
\end{center}
\end{figure}

 The specific feature of the coherent  photoproduction 
in the symmetric UPC at the collider is that there are two contributions since
both colliding proton can be sources of photons and targets. Hence, 
$J/\psi$  at fixed rapidity can be produced 
in the interaction of the low energy photon with the large $x$ gluon and in the interaction of
the high energy photon with the low $x$ gluon when source of photons and target
are interchanged. Then the cross section can be
written in the form
 \begin{equation}
{\frac {d\sigma_{pp\to ppJ/\psi}} {dtdy}}=
{\frac {dN_{\gamma /p} (y)} {dy}} 
\cdot {\frac {d\sigma_{\gamma p\to ppJ/\psi}(y,t)} {dt}}+
{\frac {dN_{\gamma /p} (-y)} {dy}} 
\cdot {\frac {d\sigma_{\gamma p\to ppJ/\psi}(-y,t)} {dt}},
\end{equation} 
where in the leading order the coherent $J/\psi$ photoproduction off the proton target can be 
described by the perturbative
QCD formula \cite{dipcoh}
\begin{equation}
{\frac {d\sigma_{\gamma p\to pJ/\psi}} {dt}}=
\frac {\Gamma _{ee} M^3_{J/\psi} \pi ^3} {48\alpha_{em}}
\cdot {\frac {\alpha_{S}^2(\bar Q^2)} {\bar Q^8}}\biggl [xg_{N} (x,\bar Q^2)\biggr ]^2
\exp[B_{J/\psi}(s)t].
\label{psicoh1}
\end{equation}  
 Here the slope $B_{J/\psi}$ is parametrized by the expression 
$$B_{J/\psi}=3.1+0.25\log_{10}(s/s_0),$$
with $s_0=100$ GeV$^2$ reasonably describing the data.

To give the prediction for the expected counting rates we calculated the coherent cross section
using  the QCD motivated formula \cite{STZ05} :
\begin{eqnarray}
 {d \sigma_{\gamma N\to J/\psi N}(s,t)\over dt}=280\cdot 
\biggl [1-\frac {(m_{J/\psi} +m_{N})^2} {s}\biggr ]^{1.5}\cdot
\biggl ({s\over 10000\,GeV^2}\biggr )^{0.415} 
\nonumber \\ 
\biggl [\Theta \bigl ({s_{0}-s}\bigr ) \biggl [ 1-{t\over t_{0}}\biggr ]^{-4} +
\Theta ({s- s_{0}}) exp(B_{J/\psi}t)\biggr ].
\label{eq:cs}
\end{eqnarray}
with free parameters fitted  
 to the existing data \cite{Adloff:2000vm}.
The scale parameter $t_0$ was fixed, $t_{0}=1 $ GeV$^2$, and the slope parameter 
for $J/\psi\,N$ scattering was 
taken to be $B_{J/\psi}(s)=4.5$ GeV$^{-2}$. Based on (Eq.\ref{eq:cs})
 prediction\cite{STZ05} of
the cross section of $J/\psi$ photoproduction
in ultraperipheral AuAu collisions at energies of RHIC  was  recently 
confirmed by experimental data
obtained by PHENIX at rapidity $y=0$ \cite{Afanasiev:2009hy}.  
Also at $\sqrt {s}=1.96$ TeV this parametrization 
gives the cross section 2.53 nb at midrapidity  which is
about 30\% lower than the CDF result $d\sigma(y=0)/dy =3.92\pm 0.25(stat)\pm 0.52(syst)$ nb
\cite {Aaltonen:2009kg}.

The calculated cross section of coherent $J/\psi$ photoproduction in ultraperipheral proton-proton
collisions  at $\sqrt{s}=7$ TeV is shown in Fig.\ref{psicoh}. 
Contrary to the $J/\psi$ photoproduction in heavy ion UPC in the proton-proton case there
is dominance of contribution to the cross section 
 from production of $J/\psi$ by the high energy photon and small $x$ gluon 
 (the dashed curve in Fig.\ref{psicoh}).  This is because
of the presence of the high energy photons  in the photon flux of the very fast moving proton
due to the rather slow drop of the proton form factor comparing to that of heavy 
ions. 
As a result contributions from the small $x$ gluons from different 
protons is reasonably 
well separated in the rapidity distribution.
If the LHC luminosity 
at $\sqrt{s}=7$ TeV will be $\approx 10^{32} cm^{-2} s^{-1}$ 
about 10$^4$  events of coherent  photoproduction of $J/\psi$ decaying into the dimuon channel
can be
 accumulated
for the 1 year running period of the ALICE detector in the interval of the $J/\psi$ rapidities
$2<y<4$ at estimated acceptance $\approx 0.05$. 
Hence, basing on the perturbative QCD analysis (Eq.\ref{psicoh1}) and 
measurement of the cross section  with $J/\psi$
rapidities in the range $2<y<4$ one will be able
to determine with reasonable accuracy
behavior of the gluon density in the proton at $x$ ranging down to $x\approx 10^{-5}$, 
i.e. in the region which
was not experimentally available so far. 
This conclusion relies on the assumption that contribution of the coherent 
photoproduction given by UPC mechanism 
 dominates the discussed cross section.
A generally accepted method to suppress / estimate the background processes 
(in the situation when the forward proton is not detected by the Roman pot detectors) 
is to use Zero Degree Calorimeters (ZDC). In the background processes the probability of proton 
breakup is high and  a large fraction of these processes ($\geq 50\%$) leads to detecting
of particles in ZDC and other forward detectors.

An example of the background 
production
mechanism discussed in the literature is 
the odderon-Pomeron interaction resulting in 
the exclusive $J/\psi$ production in pp collisions. It   
was recently considered for the  Tevatron and the  LHC  energies \cite{Bzdak:2007cz}. 
We note that, in spite of 
numerous efforts, no experimental evidence for existence of the odderon was  found so far. 
The odderon-Pomeron interaction in pp collisions is a strong interaction process 
characterized by much smaller
impact parameters than those in the UPC photoproduction. 
Within the model used in \cite{Bzdak:2007cz} the suppression factor,
the odderon/photon ratio of $J/\psi$ production cross sections integrated 
over the transverse 
momenta $p_t$ of the outgoing protons, is proportional to ${\alpha_s}^3 \cdot S^2$.
Under the assumption that the strong interaction coupling constant $\alpha_s =0.75$ and 
the gap survival probability $S^2 (LHC) =0.03$
the authors of \cite{Bzdak:2007cz} estimate this suppression factor to be $\sim 0.1$.
Comparing these
values of $\alpha_s$ and $S^2$ 
 to  $\alpha_S\sim 0.2\div 0.3$ used in the charmonium phenomenology 
and the currently accepted value of  $S^2(LHC)\le 1.5 \%$ one could reasonably expect that
the contribution of the odderon-Pomeron mechanism 
will constitute at most a few percent correction in the kinematics we consider in the paper.
 As estimated in  \cite{Bzdak:2007cz}  with an increase of $p_t$, say to
$p_{t}\approx 0.4\div 0.5$ GeV/c,
 the odderon exchange $J/\psi$ production cross 
section drops by 1 order comparing
 to the value at small $p_t$ but the photon induced mechanism at large $p_t$ 
is suppressed more significantly.
 Hence, the selection of the large transverse momentum of the outgoing protons could give one the chance
 to look for revealing of the odderon in the considered processes of exclusive $J/\psi$ production, 
  but this definitely will not be possible in the 2010 LHC run. 
The odderon exchange mechanism is also enhanced in a case when the proton in the proton-odderon vertex
dissociates into hadrons since the photon exchange inelastic 
transitions $p\to M_X$ comparing to the elastic one 
are suppressed 
by a  factor $p_t^2$ due to the gauge invariance. This contribution can be experimentally
rejected by a veto from ZDC and other forward multiplicity detectors 
installed in ATLAS, CMS and ALICE.
 Of course, in studies of the large $t$ and rapidity gap $J/\psi$ photoproduction with
 dissociation of a target proton one has to apply such a veto to the emission in one direction only.
Note also that odderon - Pomeron 
interactions can lead only to production of isoscalar ($I=0$) states. Hence it   does not contribute
 to the production of $\rho^0$-mesons and the  comparison of the 
  $\rho/J/\psi $ ratio in the large $t$ kinematics  at the LHC and at HERA could 
help to probe contribution of the odderon.

 In conclusion, 
 we would like to emphasize that experimental measurement
 of the coherent and the large momentum transfer
and rapidity gap  photoproduction of vector mesons in ultraperipheral proton-proton collisions
at $\sqrt{s}=7$ TeV during the first year of the LHC operation opens new opportunities for
the study of small x physics.

We would like to thank L~Frankfurt  for useful discussions.
This work was supported in part by the US DOE Contract Number
DE-FG02-93ER40771 and by the Program of Fundamental Researches at LHC
of the Russian Academy of Science.

\end{document}